\UseRawInputEncoding
\documentclass[twocolumn]
{aastex631}

\usepackage{bm} 
\usepackage{enumitem}
\usepackage{amsmath}
\usepackage{rotating}
\usepackage{breqn}

\newcommand\edits[1]{#1}

\makeatletter
\DeclareRobustCommand{\em}{%
  \@nomath\em \if b\expandafter\@car\f@series\@nil
   ormalfont \else \bfseries \fi}
\makeatother

\begin{document}

\title{HETDEX: Star Formation Stochasticity Diagram of Lyman Alpha Emitting Galaxies at Cosmic Noon Confirms Three Archetypes}


\author[0000-0002-9811-2443]{Nicole M. Firestone}
\affiliation{Department of Physics and Astronomy, Rutgers, the State University of New Jersey, Piscataway, NJ 08854, USA}

\author[0000-0003-1530-8713]{Eric Gawiser}
\affiliation{Department of Physics and Astronomy, Rutgers, the State University of New Jersey, Piscataway, NJ 08854, USA}
\affiliation{School of Natural Sciences, Institute for Advanced Study, Princeton, NJ 08540, USA}

\author[0000-0001-9298-3523]{Kartheik G. Iyer}
\affiliation{Columbia Astrophysics Laboratory, Columbia University, 550 West 120th Street, New York, NY 10027, USA}

\author[0000-0002-8085-7578]{Charlotte Olsen}
\affiliation{Department of Physics, New York City College of Technology, Brooklyn, NY 11201, USA}

\author[0000-0001-8525-4920]{Antara Basu-Zych}
\affiliation{Department of Physics, University of Maryland Baltimore County, Baltimore, MD 21250, USA}
\affiliation{NASA Goddard Space Flight Center, Code 662, Greenbelt, MD 20771, USA}
\affiliation{Center for Research and Exploration in Space Science and Technology, NASA/GSFC, Greenbelt, MD 20771, USA}

\author[0000-0002-7767-5044]{Adam Broussard}
\affiliation{Department of Physics and Astronomy, Rutgers, the State University of New Jersey, Piscataway, NJ 08854, USA}

\author{Peter Kurczynski}
\affiliation{NASA Goddard Space Flight Center, Greenbelt, MD 20771, USA}

\author[0000-0003-3004-9596]{Kyoung-Soo Lee}
\affiliation{Department of Physics and Astronomy, Purdue University, 525 Northwestern Ave., West Lafayette, IN 47906, USA}

\author[0000-0002-2307-0146]{Erin {Mentuch Cooper}}
\affiliation{Department of Astronomy, The University of Texas at Austin, 2515 Speedway Boulevard, Austin, TX 78712, USA}

\author[0000-0002-1328-0211]{Robin Ciardullo}
\affiliation{Department of Astronomy \& Astrophysics, The Pennsylvania
State University, University Park, PA 16802, USA}
\affiliation{Institute for Gravitation and the Cosmos, The Pennsylvania
State University, University Park, PA 16802, USA}

\author[0000-0003-2575-0652]{Daniel J. Farrow}
\affiliation{E. A. Milne Centre for Astrophysics
University of Hull, Cottingham Road, Hull, HU6 7RX, UK}
\affiliation{Centre of Excellence for Data Science,
Artificial Intelligence \& Modelling (DAIM),
University of Hull, Cottingham Road, Hull, HU6 7RX, UK}

\author[0000-0001-8519-1130]{Steven L. Finkelstein}
\affiliation{Department of Astronomy, The University of Texas at Austin, Austin, TX 78712, USA}
\affiliation{Cosmic Frontier Center, The University of Texas at Austin, Austin, TX 78712, USA}

\author[0000-0001-6842-2371]{Caryl Gronwall}
\affiliation{Department of Astronomy \& Astrophysics, The Pennsylvania
State University, University Park, PA 16802, USA}
\affiliation{Institute for Gravitation and the Cosmos, The Pennsylvania State University, University Park, PA 16802, USA}

\author[0000-0003-3428-7612]{Ho Seong Hwang}
\affiliation{Department of Physics and Astronomy, Seoul National University, 1 Gwanak-ro, Gwanak-gu, Seoul 08826, Republic of Korea}
\affiliation{SNU Astronomy Research Center, Seoul National University, 1 Gwanak-ro, Gwanak-gu, Seoul 08826, Republic of Korea}
\affiliation{Australian Astronomical Optics - Macquarie University, 105 Delhi Road, North Ryde, NSW 2113, Australia}

\author[0009-0008-4022-3870]{Byeongha Moon}
\affiliation{Korea Astronomy and Space Science Institute, 776 Daedeokdae-ro, Yuseong-gu, Daejeon 34055, Republic of Korea}

\author[0000-0002-6186-5476]{Shun Saito}
\affiliation{Institute for Multi-messenger Astrophysics and Cosmology, Department of Physics, Missouri University of Science and Technology, 1315 N. Pine St., Rolla MO 65409, USA}
\affiliation{Kavli Institute for the Physics and Mathematics of the Universe (WPI), The University of Tokyo Institutes for Advanced Study (UTIAS), The University of Tokyo, Chiba 277-8583, Japan}

\author[0000-0001-7240-7449]{Donald P. Schneider}
\affiliation{Department of Astronomy \& Astrophysics, The Pennsylvania State University, University Park, PA 16802, USA}
\affiliation{Institute for Gravitation and the Cosmos, The Pennsylvania State University, University Park, PA 16802, USA}

\begin{abstract}

In this work, we aim to measure the star formation stochasticity of Lyman Alpha Emitting Galaxies (LAEs) at Cosmic Noon. We identify 270 LAEs from the HETDEX Survey in the COSMOS field with rest-UV-through-NIR photometry from CANDELS\null. For each LAE, we perform non-parametric gaussian-process star formation history (SFH) reconstruction using the Dense Basis method. Our HETDEX LAE sample is described well by the three SFH archetypes defined for ODIN LAEs in \citet{firestone25} with comparable frequency: \textit{First Burst}, \textit{Dominant Burst}, and \textit{Nondominant Burst}. The rapidly rising Star Formation Rates (SFRs) of First Burst LAEs are not adequately represented in traditional SFR$_{100}$-M$_*$ diagrams, where SFR$_{100}$ is averaged over the most recent 100\,Myr. This motivates the usage of SFR$_{10}$-M$_*$, where SFR$_{10}$ is averaged over the most recent 10\,Myr. We introduce the Star Formation Stochasticity Diagram, a diagnostic tool that probes variations in galaxies' SFRs across cosmic time. By \edits{mitigating} the confounding factor of redshift evolution, we are able to employ a gaussian mixture model to decompose our ratio of short- vs.\ long-term SFR into statistically motivated sub-populations. This \edits{independent} component decomposition reveals three populations consistent with the empirical LAE SFH archetypes. We can, therefore, summarize the overall stellar mass assembly of LAEs with this ratio, even without viewing their full star formation histories.

\end{abstract}



\section{Introduction}\label{sec:intro}


Lyman Alpha Emitting (LAE) galaxies are characterized by their luminous (rest-frame) ultraviolet emission, which signals the presence of O- and B-type stars at the time of observation \citep[e.g.,][]{kunth1998hst, hui1997equation}. As this ultraviolet light travels from a distant LAE to the observer, it is redshifted according to the galaxy's precise lookback time. Because of this, LAEs act as signposts in the observed optical band, illuminating the high-$z$ \edits{U}niverse. During Cosmic Noon ($2\leq z \leq 3$), LAEs are often used as statistical probes of galaxy formation and evolution \citep[e.g.,][]{odin_survey} as well as cosmology \citep[e.g.,][]{adams2010hetdex, schlegel2022spectroscopic}. Higher redshift LAEs ($6\leq z \leq 14$) are powerful tools for investigating the reionization era of the \edits{U}niverse \citep[e.g.,][]{kikuta2023silverrush}. To draw an evolutionary connection between these studies, one needs to understand the stage of galaxy evolution that LAEs are in. Several studies have shown that, while most LAEs are young, low-mass, and low-dust objects, LAEs samples often also include sub-populations of higher mass galaxies \citep[e.g.][]{gawiser2007lyalpha, finkelstein2009lyman, guaita2011lyalpha, vargas2014stack, acquaviva2012curious, kusakabe2018stellar}. The presence of high mass LAEs is particularly compelling, as the detection of Ly$\alpha$ in galaxies necessitates both strong intrinsic Ly$\alpha$ emission and low propensity for absorption by dust in the interstellar medium (ISM) and intergalactic medium (IGM\null). This scenario is easily explainable in young, star-bursting, low-mass galaxies that have yet to produce significant amounts of dust through prolonged star formation. However, to explain high mass LAEs, one needs a more nuanced understanding of stellar mass assembly and radiative transfer. In brief, the results to date suggest that LAEs have complex, non-uniform evolutionary pathways and we currently lack an understanding of the full picture of their stellar mass assembly and radiative transfer\citep[see][]{rosani2020bright, iani2024midis, ceban2024star, firestone25}. 

One key tool that is often used to compare the star forming phase of LAEs with other galaxies is the SFR-M$_*$ diagram, correlating a galaxy's total stellar mass at the time of observation ($M_*$) to its Star Formation Rate (SFR). Several studies \citep[e.g.,][]{brinchmann2004physical, noeske2007star, daddi2007multiwavelength, elbaz2007reversal} report that generic star forming galaxies trace a fundamental relationship on the SFR-M$_*$ diagram. This relationship is often referred to as the SFR-M$_*$ \textit{correlation} or the \textit{star forming main sequence}. Above the correlation, we expect to find star-bursting galaxies; below the correlation, we expect to find quiescent galaxies. Therefore, the SFR-M$_*$ correlation serves as a useful benchmark to compare the star forming behavior of LAEs to other galaxies. 

\citet{vargas2014stack} performed an SFR-M$_*$ analysis of a sample of $z = 2.1$ photometrically selected LAEs from the Multi-wavelength Survey by Yale-Chile \citep[MUSYC;][]{gawiser2006multiwavelength} with photometry from the Cosmic Assembly Near-Infrared Deep Extragalactic Legacy Survey \citep[CANDELS;][]{candels1, candels2}. Their LAEs fell above the SFR-M$_*$ correlation of \citet{daddi2007multiwavelength} derived from massive galaxies at $1.5 < z < 2.5$ \citep[see Figure 8 in][]{vargas2014stack}. Moreover, their measured scatter in SFR-M$_*$ was most pronounced at low mass, suggesting that these galaxies are experiencing ongoing starbursts.  Similarly, \citet{hagen2014spectral} performed an SFR-M$_*$ analysis of a sample of spectroscopically selected LAEs from the Hobby-Eberly Telescope at $1.9 < z < 3.6$. These LAEs also resided above the SFR-M$_*$ correlation of \citet{daddi2007multiwavelength}, but along the main sequences of \citet{whitaker2012star} for $z$ = 2.0 and $z$ = 3.5 galaxies \citep[see Figure 10 in][]{hagen2014spectral}. \citet{finkelstein2015probing} utilized a sample of $z = 4.5$ LAEs identified with Spitzer and Hubble Space Telescope data; their LAEs tend to lie above the correlations of \citet{daddi2007multiwavelength}, \citet{hathi2013stellar}, and \citet{speagle2014highly}, except for the most massive LAEs which lie on the  \citet{hathi2013stellar} correlation \citep[see Figure 22 in][]{finkelstein2015probing}. \citet{shimakawa2017similarities} utilized a sample of $z = 2.5$ LAEs with Subaru Prime Focus Camera data; their LAEs lay along the correlations derived from \citet{whitaker2014constraining, shivaei2015mosdef, tomczak2016sfr} \citep[see Figure 8 in][]{shimakawa2017similarities}. \citet{kusakabe2018stellar} reported that a sample of $z = 2$ LAEs lie along the correlations derived from \citet{daddi2007multiwavelength}, \citet{tomczak2016sfr}, and \citet{shivaei2017mosdef} \citep[see Figure 15 in][]{kusakabe2018stellar}. While these results deepen our understanding of the relative star forming behavior of LAEs, they highlight a lack of consensus among the literature. This uncertainty may be driven by several fundamental challenges in $SFR-M_*$ analyses. 

To self-consistently compare the behavior of galaxies identified through various selection techniques, the same base assumptions must be made across samples i.e., SED fitting must be conducted using a consistent initial mass function (IMF), dust law, stellar isochrones, spectral libraries, and fitting algorithm. Differences in these base assumptions have led to significant disagreements in star formation rate and stellar mass measurements across various surveys \citep[e.g.][]{kennicutt2012star, speagle2014highly, hagen2014spectral, finkelstein2015probing, shimakawa2017similarities}. These results caution against direct comparisons between star formation rates and stellar masses computed using different algorithms and priors. 

As of this publication, the largest methodological variation across the literature involves the spectral energy distribution (SED) fitting and/or star formation history (SFH) reconstruction algorithm. This aspect is particularly important when working with star-bursting galaxies like LAEs, which are very sensitive to SFH parameterization. It is, therefore, paramount to use an SFH reconstruction algorithm that is able to produce realistic, continuous SFHs that are not restricted by functional form. 
An additional challenge with SFR-M$_*$ analyses is that the relationship between SFR and $M_*$ is known to evolve with time \citep[see][]{daddi2007multiwavelength}; this behavior reduces the statistical power of datasets that span wide ranges of redshifts, and introduces systematic scatter in the correlation. While some studies have remedied this issue by regressing across cosmic time \citep[e.g.][]{speagle2014highly, boogaard2018muse, goovaerts2024galaxy}, this approach requires assumptions about the parameterization of the evolution. This complication motivates a search for new techniques to robustly assess the stochasticity of LAE star formation compared to that of other star forming galaxies. In addition to the direct challenges introduced by the methodology of measuring the SFR-M$_*$ correlation, the suggestion that the behavior of LAEs with respect to the correlation may vary based on stellar mass \citep{finkelstein2015probing} requires us to better understand how the diversity of LAE properties may impact our interpretation of their star forming behavior. To address this issue and the deficiencies above, one must perform a self-consistent analysis that acknowledges the diversity in LAE stellar mass assembly, by enabling detailed, realistic, and flexible star formation history reconstruction.

\edits{In addition to SFR-M$_*$, another diagnostic for measuring the star forming phase of galaxies is often refereed to as \textit{star formation stochasticity} or \textit{burstiness}, whereby we compare short-term and long-term star formation across a population of galaxies. \citet{broussard2019star, broussard2022improved} define the theoretical burst indicator as $\eta_{th} = \log_{10}(\text{SFR}_{10}/\widetilde{\text{SFR}}_{100})$, where the width of this distribution demonstrates the stochasticity for a population of galaxies. This definition is comparable to those presented in \citet{weisz2012modeling}, \citet{guo2016bursty}, and references therein. Observationally, this burst indicator can be measured when information is available from H$\alpha$ and the UV continuum, which are proxies recent and long-term SFR, respectively: $\eta = \log_{10}(\text{SFR}_{\text{H}\alpha}/{\text{SFR}}_{\text{NUV}})$ \citep[][]{broussard2019star}. Similar measurements have been made for galaxies across redshifts \citep[e.g.][]{faisst2019recent, mehta2023spatially}. }

In this work, we build on the analysis of \citet{firestone25}, which utilized nonparametric star formation histories of LAEs photometrically selected from the One-hundred-deg$^2$ DECam Imaging in Narrowbands survey \citep[ODIN][]{odin_survey}. From these star formation histories, \citet{firestone25} identified three LAE stellar mass assembly scenarios, or archetypes (see Section \ref{subsec:archetypes}), with $\sim95\%$ of LAEs experiencing a ``formative burst'' of star formation at the time of observation. In this paper, we utilize a spectroscopically confirmed LAE sample from the Hobby-Eberly Telescope Dark Energy Experiment \citep[HETDEX][]{gebhardt2021hobby}. We cross-match the HETDEX LAE sample with the CANDELS photometric catalog in the COSMOS field to create sub-samples of LAEs with rest-frame UV-through-NIR photometry. With these data, we implement the Dense Basis approach to SED fitting and star formation history reconstruction \citep{iyer2017reconstruction} and address several key pitfalls in the SFR-M$_*$ diagram. We introduce new techniques for measuring the star formation stochasticity of extreme star-bursting galaxies across cosmic time. Throughout this analysis, we compare the behavior of our LAEs to the \citet{merida2026probing} catalog of generic star forming galaxies analyzed using the same base assumptions as our LAE catalog. Section~\ref{sec:data} introduces the HETDEX's LAE sample and the CANDELS photometric catalog. Section~\ref{sec:SFHs} presents the Dense Basis priors used in this analysis and discusses the results of the star formation history reconstruction in the context of the approach taken by \citet{firestone25}. Section~\ref{sec:sfr_mstar} describes the relationship between star formation rate and stellar mass, compares our results to the generic star forming galaxy correlation from \citet{merida2026probing} and discusses the limitations of the SFR-M$_*$ diagram. Section~\ref{sec:stochasticity} introduces the Star Formation Stochasticity Diagram and explores new techniques for qualifying stellar mass assembly phase in galaxies. We then apply this technique to our sample of LAEs. The major conclusions of this study are summarized in Section~\ref{sec:conc}. Throughout this work, we assume a $\Lambda$CDM cosmology with $h$ = 0.7, $\Omega_m$ = 0.27, and $\Omega_{\Lambda} = 0.73$.

\section{Data}\label{sec:data}


\subsection{HETDEX LAE Catalog}\label{subsec:HETDEX_LAEs}



HETDEX is a untargeted spectroscopic survey designed to map out the large scale structure of LAEs of the $1.9 < z < 3.5$ \edits{U}niverse \citep{gebhardt2021hobby, mentuch2026hetdex}. To do the experiment, HETDEX utilizes the Visible Integral-field Replicable Unit Spectrograph (VIRUS), a set of 78 $50\arcsec \times 50\arcsec$ integral-field units (IFUs) distributed across the focal plane of the Hobby-Eberly Telescope \citep{hill2021hetdex}, at the McDonald Observatory at The University of Texas.  Since each spectrograph feeds two low resolution ($R \sim 800$) spectrographs covering the wavelength range $3500 < \lambda < 5500$~\AA, virtually all the LAEs detected are identified by a single line of Ly$\alpha$. Ancillary deep imaging from a variety of sources helps discriminate true LAEs from foreground contaminants with better 98\% accuracy \citep{davis2023hetdex}.

The primary goal of HETDEX is to understand how dark energy fuels the expansion of the \edits{U}niverse across cosmic time. To accomplish this task, HETDEX uses LAEs as tracer of the underlying matter distribution through their dark matter halos. This goal necessitates a fundamental understanding of the galaxy-halo connection (i.e., the halo occupation distribution) for LAEs. One approach to probe this connection is to quantify the uniformity (and diversity) of LAE properties, notably their star formation stochasticity. 

For this analysis we cross match HETDEX Public Source Catalog 2 (HPSC2), the catalog released with the HETDEX Public Data Release 1 (PDR1, \citealt{mentuch2026hetdex}. HPSC2 is composed of astronomical sources gathered from 8 years of HETDEX main survey data. The catalog generation including source detection, detection grouping, redshift assignment and classification is described in detail in HETDEX Public Source Catalog 1 \citep{mentuch2023hetdex} and updated in \citet{mentuch2026hetdex}. We summarize the relevant selection criteria described in \citep{mentuch2023hetdex} and \citet{mentuch2026hetdex} below. 

Firstly, potential LAEs are identified by performing a grid search for emission line features using the techniques described in \citet{gebhardt2021hobby}. The continuum emission is then subtracted and the signal-to-noise ratio and $\chi^2$ of the fit are computed. The location of the object is defined as the position that maximizes signal-to-noise ratio. Each of the LAEs has a continuum subtracted signal-to-noise ratio of at least 4.8 and an emission line gaussian fit $\chi^2$ of less than 2.5. Additionally, a Gaussian line width between 1.6 and 14\,\AA{ }is required to eliminate objects with calibration issues. Then, several data quality cuts are applied to mitigate contamination from detector artifacts due to poor calibrations, which can result in negative continuum flux measurements \citep{mentuch2023hetdex}. To address this issue, the continuum flux density is required to exceed $-3 \times 10^{-17}$ erg s$^{-1}$ cm$^{-2}$ \AA$^{-1}$. Additionally, the highest weighted fiber profile fit must have a quality $\chi^{2}_{fiber}$ of less than 4.5. Lastly, to account for drops in transparency due to cloud coverage, the relative throughput of the observation (scaled to a 360s observation) must exceed 0.08. 

\subsection{CANDELS Photometric Catalog}\label{subsec:CANDELS}

The Cosmic Assembly Near-IR Deep Extragalactic Legacy Survey \citep[CANDELS;][]{candels1, candels2} is a photometric imaging survey aimed at probing the galaxy population during Cosmic Noon (and Dawn) using the \textit{Hubble Space Telescope}'s near-infrared Wide Field Camera 3 (WFC3) and optical Advanced Camera for Surveys (ACS\null). 
Together with photometric data from CFHT \citep{cfht}, Subaru \citep{subaru1, subaru2}, {Ultra}VISTA \citep{ultravista}, IRAC \citep{irac_seds, s-cosmos}, and NEWFIRM \citep{newfirm}, CANDELS provides photometric measurements from 46 filters spanning a wavelength range of roughly 250-9340\,nm \citep[see Table 1 of][]{firestone25}. 


\section{Star Formation Histories}\label{sec:SFHs}


\subsection{SED Fitting and SFH Reconstruction}\label{subsec:SED_fitting}



A galaxy's spectral energy distribution can be decomposed into several key components: stellar continuum, nebular emission, dust absorption and emission, and (when present) AGN emission. These components, respectively, can be used to derive the galaxy's present and past star formation, ISM conditions, dust properties, and obscured star formation. Together, these data can be used to synthesize the star formation history of a galaxy -- an archaeological record of a galaxy's formation and evolution across its cosmic lifetime. For this analysis, we implement the Dense Basis technique of \citet{db1-iyer2017} and \citet{db2-iyer2019}. The Dense Basis approach is a non-parametric Gaussian processes-based method for SED fitting and star formation history reconstruction \citep[see also][]{MOPED, MOPED_SFH, VESPA_SFH, prospector_leja2017deriving, prospector_johnson2021stellar}. 
The Gaussian process element of Dense Basis allows reconstruction of smooth star formation histories with non-specific functional forms; this approach minimizes bias in SFH parametrization at all lookback times and allows for star formation histories with several peaks. 

For galaxies experiencing active star formation, parametric SFHs restricted to single peaks are highly likely to conclude that this peak occurred at the time of observation. This parameterization introduces significant biases in SFH solutions by effectively prohibiting significant past star formation \citep[see][]{simha2014parametrising, carnall2019measure, dirichlet2}. \edits{This effect is often referred to as \textit{outshining}, and is particularly relevant when we are aiming to better constrain that recent star formation in the context of a galaxy's older stellar populations.} 

Dense Basis is not restricted by any parametric model, allowing for multiple episodes of star formation spanning the lifetime of the galaxy. Simulations have shown that Dense Basis produces robust and detailed results that align with a galaxy's true properties, with a scatter of 0.2 dex out to a lookback time of $\sim$5 Gyr \citep{db2-iyer2019}. This duration is sufficient for our sample, which reaches a maximum lookback time of $\sim$\edits{3.6}\,Gyrs. These advantages make Dense Basis the ideal tool for accurately assessing the diverse stellar mass assembly of LAEs while minimizing systematic limitations. \edits{That being said, there are also notable caveats to consider when working with non-parametric SFHs. Firstly, non-parametric SFH reconstruction tends to be computationally intense. However, Dense Basis addresses this through the utilization of an ``atlas'' containing a \textit{dense basis} of possible SFHs drawn from a carefully designed set of broad priors and physically motivated stellar isochrones. This feature significantly reduces computational expense. Secondly, since non-parametric SFH reconstruction is designed to be flexible, results can be heavily influenced by priors. We combat this with intentional choices of priors. These choices and their impact on the sample are discussed below.}

For our SED fitting and SFH reconstruction, we adopt priors that are broadly consistent with the observed properties of LAEs without heavily restricting to our assumptions \citep[e.g.,][]{gawiser2007lyalpha, guaita2011lyalpha, vargas2014stack, acquaviva2012curious, firestone25}. We curate a diverse SFH prior space that allows for extreme star-bursting and steady star forming solutions in order to avoid introducing significant bias in the output. 
Following the priors of \citet{firestone25}, we assume the metallicity ($Z$) range of log($Z$/$Z_\odot$) is uniform from $-$1.5 to 0.2, the prior for the lookback times ${t_x}$ is Dirichlet \edits{(multivariate generalization of the beta distribution)} with $\alpha$ = 3.0 \citep{prospector_leja2017deriving, dirichlet2}\edits{, and we use 3 ${t_x}$ parameters: ${t_{25}}$, ${t_{50}}$, and ${t_{75}}$, which represent the times at witch \textit{x}\% of the total stellar mass was formed}. We also assume a Chabrier IMF \citep{chabrier2003galactic} and a Calzetti dust model \citep{calzetti2000dust} with a flat dust extinction $A_V$ prior from 0.0 to 2.0. Lastly, we assume a flat redshift distribution, centered at each galaxy's HETDEX spectroscopic redshift with a width of 0.05. \edits{As was discussed in \citet{firestone25}, we find that our posterior distributions for stellar mass and star formation rate do not push up against the bounds of the priors, indicating that our priors do not introduce significant bias in our measurements. We also find that changing the number of ${t_x}$ parameters used in the priors does not produce any notable difference in the SFH classification or number of major SF bursts in the samples. 
These findings suggest that, despite the use of non-parametric SFHs, our results do not suffer from significant biases due to prior choices.}

To perform our SED fitting, we begin with all CANDELS filters \citep[see Table 1 of][]{firestone25} except for IRAC channels 3 and 4, which suffer from large point spread functions and photometric errors. To account for incomplete modeling of complex Ly$\alpha$ radiative transfer (which is pervasive in stellar population synthesis models) for each galaxy, we exclude any intermediate band filters that contain the Ly$\alpha$ emission line. \edits{Even though we exclude intermediate band filters that contain the Ly$\alpha$ emission line, we still retain sensitivity to 10\,Myr timescales due to our inclusion of nebular emission lines (e.g. H$\alpha$, H$\beta$, \ion{O}{3}) in the rest-frame optical regime.} We also use observed photometric error floors from \citet{firestone25} to account for systematic errors and SED template incompleteness \citep{iyer2019reconstructing}.

\subsection{LAE Sample}\label{subsec:lae_sample}

To produce the LAE sample for this analysis, we positionally cross match the HETDEX LAE catalog with the CANDELS catalog (requiring a positional agreement within 1 arcsecond), utilizing CANDELS photometry. We then require that each object in our sample has a Dense Basis reduced $\chi^2$ values $\leq$ 2.6 \citep[based on][]{firestone25}. Although we have spectroscopic data for our HETDEX LAE sample, this step is useful for ensuring high quality SED solutions and further limits contamination from spurious detections, interloping galaxies (namely [\ion{O}{2}] emitters), and AGN that also produce Ly$\alpha$ emission. After the reduced $\chi^2$ cut, we retain $\sim$\,61$\%$ \edits{of the initial 444 objects}, resulting in a final sample of 270 LAEs. \edits{The remainder of this analysis is conducted on the resultant sample of 270 LAEs with high SED fit quality.}




\subsection{SFH Archetypes}\label{subsec:archetypes}


In this work, we distinguish LAEs based on their star formation history archetypes, as defined in \citet{firestone25}. These archetypes are used to delineate between the different stellar mass assembly pathways that may produce an LAE. The three archetypes are summarized below: 

\begin{itemize}

    \item \textbf{First Burst}: galaxies undergoing their first burst of star formation in the last 200\,Myr, with at most modest star formation (SFR $< 1$\,M$_\odot \text{yr}^{-1}$) more than 200\,Myr ago. 

    \item \textbf{Dominant Burst}: galaxies undergoing their largest burst of star formation in the last 200\,Myr, but with significant star formation (SFR $\geq 1$\,M$_\odot \text{yr}^{-1}$) in the past. 

    \item \textbf{Nondominant Burst}: galaxies where the largest burst of star formation occurred in the past (over 200\,Myr ago). 
    
\end{itemize}
The \textit{First Burst} archetype and \textit{Dominant Burst} archetype together characterize LAEs that are undergoing a \textit{formative} burst of star formation, i.e., these LAEs are experiencing a starburst that is paramount in the formation of the galaxy \citep{firestone25}. 

\subsection{SFH Results}\label{subsec:SFH_results}

\citet{firestone25} reports that 67\%, 28\%, and 5\% of 74 photometrically-selected ODIN LAEs in the COSMOS field fell into the First Burst, Dominant Burst, and Nondominant Burst archetypes, respectively. Additionally, the Formative Burst characterization is consistent with $\sim95\%$ of the total ODIN LAE sample. In this work, 61\%, 31\%, and 7\% of 270 spectroscopically-selected HETDEX LAEs fall into the First Burst, Dominant Burst, and Nondominant Burst archetypes, and the Formative Burst characterization is consistent with $\sim92\%$ of the total HETDEX LAE sample. These values are consistent with Figure 1 of \citet{firestone25}. The results from HETDEX and ODIN are remarkably similar, indicating a robust trend that spans a broad redshift range. 


\section{Star Formation Rate vs. Stellar Mass}\label{sec:sfr_mstar}

\subsection{$SFR-M_*$ Methodology}\label{subsec:sfr-m*_methodology}

Typical characterizations of galaxy star forming phase rely on the SFR-M$_*$ diagram, comparing the total stellar mass at the time of observation ($M_*$) and the Star Formation Rate (SFR) \citep[e.g.,][]{kennicutt2012star, speagle2014highly, hagen2014spectral, finkelstein2015probing, shimakawa2017similarities}. Typically, SFR-M$_*$ analyses use the star formation rate obtained from the UV continuum (SFR$_{\rm UV}$) or average the star formation rate over the last 100\,Myr (SFR$_{100}$) using SED fitting. These methods are roughly consistent since the UV continuum is driven by O- and B-type stars with ages of approximately 5-10\,Myr and 100\,Myr, respectively.

For our analysis, we use the time averaged star formation rate and the total stellar mass formed ($M_{\rm formed}$) from our star formation history reconstructions. To convert from $M_{\rm formed}$ to $M_*$, we apply a mass loss function derived from \texttt{Flexible Stellar Population Synthesis} \citep[FSPS;][]{FSPS1, FSPS2} assuming the MILES+MIST stellar isochrones \citep{MILES, MIST_2016_Choi}. Since the relationship between SFR and $M_*$ is redshift dependent \citep{daddi2007multiwavelength}, we divide our sample into three redshift bins of width $\Delta z \sim 0.5$. 

For comparison between our LAEs and generic star forming galaxies, we utilize SFR-M$_*$ correlations from \citet{merida2026probing}. These relations are derived from a combined sample of star-forming galaxies 
in COSMOS2020, CANDELS, SHARDS/CANDELS FAINT, CANUCS, CEERS, JADES, and UNCOVER \citep[][and references therein]{merida2026probing}. This extensive dataset allowed the authors to probe the star forming sequence of 755,196 galaxies from the low-$M_*$ to high-$M_*$ regime across a redshift range that is inclusive of our sample. More importantly, \citet{merida2026probing} derive their star formation rate and stellar masses values using Dense Basis with same (Chabrier) IMF and (Calzetti) dust law used in our study, thus allowing a direct comparison of their results to our LAE sample. The only notable difference between the two analyses is that \citet{merida2026probing} derive galaxy properties from the Dense Basis SED fits, while we pull galaxy properties from the Dense Basis SFHs. We adopted this approach because the SFH results better capture the star forming behavior of bursting galaxies. To account for this difference, we apply a simple \edits{upward correction of $\sim$0.2 solar masses per year} to the correlation's normalization based on the behavior of a subset of our LAEs that behave most similarly to the generic star forming galaxies located along the correlation. 

For their analysis, \citet{merida2026probing} assume a linear parameterization of the SFR-M$_*$ correlation, 
\begin{equation}\label{eq:corr}
    \log{({\rm SFR}_{100})} = \beta \log{(M_*)} + \alpha,
\end{equation} 
with symmetrical scatter in $\log{\rm SFR}$ and $\log{M_*}$. They then fit the data over their entire stellar mass range \edits{( $10^6 \lesssim M_*/M_\odot 
\lesssim 10^{11}$)} over various redshift bins using \texttt{lmfit} \citep{newville2014lmfit}, a non-linear Levenberg-Marquardt least squares algorithm. Values for the normalization $\alpha$, slope $\beta$, and the scatter $\epsilon$, for each redshift bin are presented in Table 3 of \citet{merida2026probing}.

\subsection{SFR-M$_*$ Results}

\begin{figure*}
\begin{center}
\includegraphics[width=0.9\textwidth]{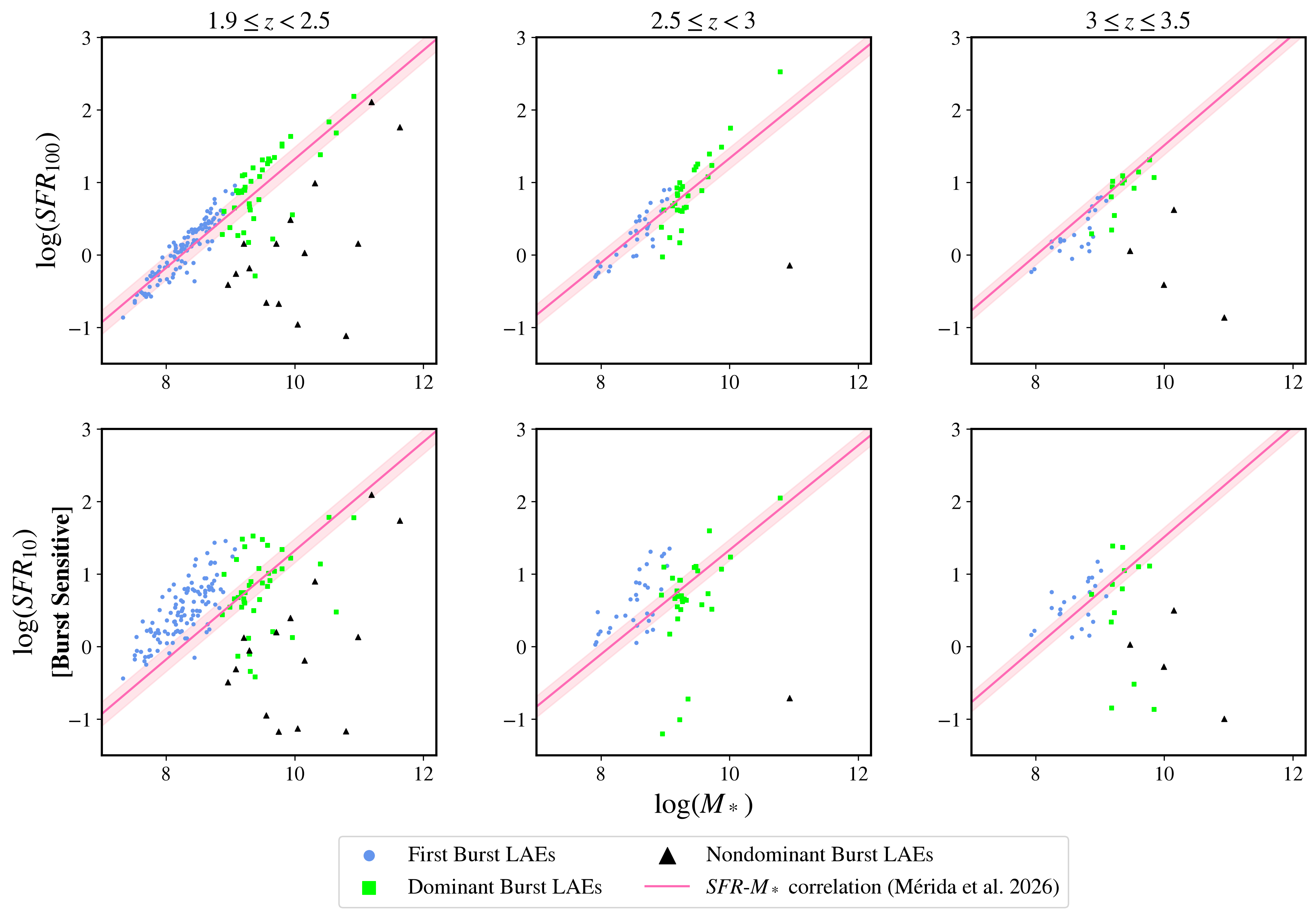}
\caption{The top row of panels presents the traditional ${\rm SFR}-M_*$ diagram while the bottom row displays a burst-sensitive version where the star formation rate is averaged over a shorter timescale. The $x$-axis represents the logarithm of the stellar mass in solar masses at the time of observation. The $y$-axis displays the logarithm of the star formation rate in solar masses per year averaged over the last 100\,Myr (top) and 10\,Myr (bottom). The left panels contain galaxies with redshifts $1.9 \leq z < 2.5$, the middle panels contain galaxies with redshifts $2.5 \leq z < 3.0$, and the right panels contain galaxies with redshifts $3.0 \leq z < 3.5$. In each panel, the galaxies are identified by their LAE star formation archetypes, with First Burst LAEs shown as blue circles, Dominant Burst LAEs indicated by green squares, and Nondominant Burst LAEs depicted as black triangles. The correlation lines in pink are from \citet{merida2026probing}. 
} 
\label{fig:sfr_m*}
\end{center}
\end{figure*}

The top panel of Figure \ref{fig:sfr_m*} presents the LAEs in SFR$_{100}$-M$_*$ space, divided into three redshift bins with $\Delta z \sim 0.5$. By examining the star formation histories
and Figure \ref{fig:sfr_m*} together, we can compare the behavior of the LAEs by SFH archetype. First Burst LAEs most notably have low stellar masses; Dominant Burst LAEs have higher stellar masses caused by a combination of later-stage burst phase and prior star formation; Nondominant Burst LAEs naively appear to be quiescent, but their star formation histories reveal that they have large amounts of past star formation that drive up $M_*$ and push them rightwards on the diagram. The behavior of Nondominant Burst LAEs is a reminder that the SFR-M$_*$ diagram is not fully deterministic of a galaxy's properties at the time of observation. 

The most notable result of the top panel of Figure \ref{fig:sfr_m*}, however, is that the star-bursting nature of the First Burst LAEs is not adequately captured, i.e., they do not appear any burstier than Dominant Burst LAEs (and sometimes look less extreme). First Burst LAEs are undergoing a very recent ramp up of star formation activity after a lifetime of virtually no star formation \citep{firestone25}. Because of this history, the ongoing burst is averaged with zero over the course of 100\,Myr in our SFR$_{100}$ measurement. This result misleadingly pulls the First Burst LAEs down in SFR$_{100}$-M$_*$ space and blends them with the generic population of star forming galaxies. To better capture the nature of young starburst galaxies, we need to use a shorter timescale star formation rate that is more sensitive to a recent/ongoing burst. We, therefore, introduce a simple ``burst sensitive'' version of the diagram utilizing the star formation rate averaged over the last 10\,Myr, SFR$_{10}$. A potential advantage of this version is that SFR$_{10}$ is more closely related to the star formation traced by emission lines, which is often used as an observational probe of short-timescale SFR\null. We present the results for the burst sensitive SFR-M$_*$ diagram in the bottom panels of Figure \ref{fig:sfr_m*}. In this version of the diagram, First Burst LAEs tend to lie above the correlation; Dominant Burst LAEs appear along the correlation; and Nondominant Burst LAEs lie below the correlation. 



\subsection{Limitations in SFR-M$_*$}\label{subsec:sfr_mstar_lims}


If the stellar mass assembly in all galaxies were governed by a uniform deterministic star formation history, there would be a single curve on the SFR-M$_*$ diagram, with the only scatter around this curve caused by observational uncertainties. Hence a key probe of the level of stochasticity in galaxies' star formation is the amount of \emph{intrinsic} scatter around the well-studied SFR-M$_*$ correlation.  

As mentioned in Section \ref{subsec:sfr-m*_methodology}, several studies of large galaxy populations have demonstrated that the SFR-M$_*$ correlation evolves with redshift. As a consequence of this evolution, SFR-M$_*$ analyses often employ redshift bins to mitigate this effect while still producing statistically robust correlations. However, if these bins are moderately wide, rapid redshift evolution in the correlation will artificially widen the distribution of points.  One way to remedy this effect is to perform a single regression in all three dimensions of $SFR$, $M_*$, and cosmic time (or redshift) \citep[e.g.][]{speagle2014highly, boogaard2018muse, goovaerts2024galaxy}. However, while powerful, this solution requires assumptions about the parameterization of both the correlation and its evolution in time, thus making it difficult to reliably measure the stochasticity of star formation. 

To begin to understand the stochasticity of star formation in LAEs and other galaxy samples, it is necessary to properly account for the evolution of star formation as a function of redshift. We, thus, introduce the Star Formation Stochasticity Diagram.

\section{The Star Formation Stochasticity Diagram}\label{sec:stochasticity}

\subsection{Star Formation Stochasticity Methodology}



We introduce a more robust version of the classic SFR-M$_*$ diagram that removes cosmic age from the evolution of the normalization, thus allowing measurement of the behavior of galactic evolution with greater precision. This new diagnostic relies on breaking $M_*$ into a product of SFR$_{\rm avg}$, $t_{\rm univ}$, and $f_{\rm ret}$, where $M_{formed}$ is the total formed stellar mass (without losses), $f_{ret}$ is the fraction of stellar mass retained after losses due to stellar death and winds, and $t_{univ}$ is the age of the Universe at the time of observation.  We then compare the galaxies' SFR to SFR$_{\rm avg}$.  To do this\edits{,} we define the long-term average star formation rate for a galaxy since the Big Bang as follows,
\begin{equation}\label{eq:sfr_avg}
    {\rm SFR}_{\rm avg} = \frac{M_{\rm formed}}{t_{\rm univ}} = \frac{M_*}{f_{\rm ret} \times t_{\rm univ}}. 
\end{equation}

\begin{figure*}
\begin{center}
\includegraphics[width=0.99\textwidth]{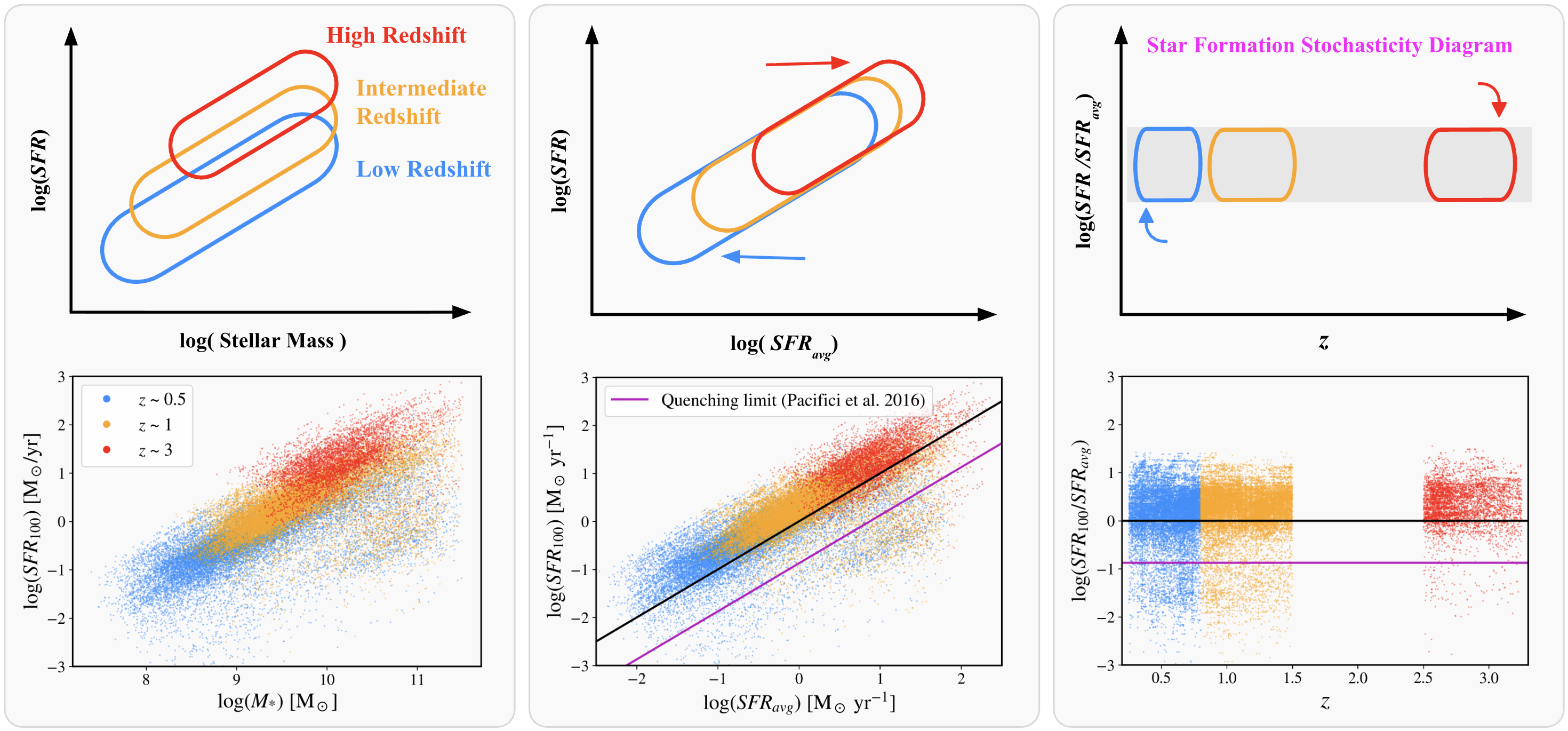}
\caption{The upper left panel is a schematic depiction of the familiar SFR-M$_*$ correlation found for star-forming galaxies at all redshifts. The upper middle panel illustrates that changing the $x$-axis to each galaxy's average SFR since the Big Bang (SFR$_{\rm avg}$) will shift low-redshift galaxies to the left and high-redshift galaxies to the right due to the different amounts of time they have had to generate a given amount of stellar mass. The upper right panel introduces the Star Formation Stochasticity Diagram. In this diagram, we shift to a $y$-axis that compares a galaxy's short-term star formation rate to its average SFR since the Big Bang (SFR/SFR$_{\rm avg}$), plotted as a function of redshift. This format rotates the data to a flattened correlation, highlighting any true evolution. As an illustration with real data, the lower left panel shows results from Dense Basis SFHs of 40,534 CANDELS galaxies in three redshift ranges, $0.2<z<0.8$ (blue points), $0.8<z<1.5$ (orange points), and $2.5<z<3.5$ (red points). The lower middle panel shows the same galaxies plotted as SFR vs. SFR$_{\rm avg}$, along with a black line showing where SFR = SFR$_{\rm avg}$ and a magenta line below which galaxies are considered to be quenched following the criterion of \citet{pacifici2016timing}; quiescent galaxies lie below this line at all three redshifts. The lower right panel shows the CANDELS galaxies on the Star Formation Stochasticity Diagram, where it becomes clear that the fraction of quenched galaxies is much higher at low redshift.}  
\label{fig:sfr_sfravg}
\end{center}
\end{figure*}
As is apparent in Figure \ref{fig:sfr_sfravg} (bottom row) for a sample of 40,534 CANDELS galaxies \citep{candels1, candels2} analyzed with Dense Basis in a similar manner to \citet{iyer2019nonparametric}, the shift from SFR-M$_*$ to SFR$-SFR_{\rm avg}$ significantly reduces redshift evolution in the correlation. This result illustrates that the SFR$-$SFR$_{\rm avg}$ correlation directly probes variations in galaxies' short-term versus long-term SFRs.

Additionally, we define the Star Formation Stochasticity Metric as $\log({\rm SFR/SFR_{avg}})$. With this metric, we introduce the \emph{Star Formation Stochasticity Diagram} (see Figure \ref{fig:sfr_sfravg}, right panels), where we examine $\log({\rm SFR/SFR_{avg}})$ as a function of redshift. This diagram allows one to empirically determine the evolution of galaxies' star formation behavior over a wide range of redshifts and to measure stochasticity for a population of galaxies via scatter in SFR/SFR$_{\rm avg}$. With this diagram, we can evaluate the stochasticity of star formation among the LAE SFH archetypes and compare those to generic star-forming galaxies across cosmic time. Additionally, the removal of redshift dependence enables us to utilize the statistics of the HETDEX sample across the entire redshift range without having to bin the data.  

Assuming a linear relationship between $\log({\rm SFR)}$ and $\log(M_*)$ \edits{(as defined in Equation \ref{eq:corr})}, we can easily shift between this metric and the classic relationship between SFR and $M_*$ by employing Equations \ref{eq:corr} and \ref{eq:sfr_avg}.
\begin{dmath}\label{eq:switch}
    \log\left(\frac{{\rm SFR}_{100}}{{\rm SFR}_{avg}}\right) = (\beta - 1)\log(M_*) + \alpha 
    \\+ \log(f_{\rm ret} \times t_{\rm univ})
\end{dmath}



The Star Formation Stochasticity Metric shows some similarities with the well-established \textit{Scalo $b$ parameter} (or \textit{birthrate parameter}) \citep{scalo1986stellar, kennicutt1998star}, which is almost exclusively implemented in the low-redshift \edits{U}niverse.
However, there are several key differences in both its computation and application. The Scalo $b$ parameter is often defined as
\begin{equation}
    b = \frac{SFR}{M_*} \times t_{gal},
\end{equation}
where SFR is typically taken as the recent star formation rate measured by tracers such as H$\alpha$, $M_*$ is the stellar mass at the time of observation, and $t_{gal}$ is the age of the galaxy defined from the formation of its first stars. 

While similar, our method utilizes the total stellar mass formed in the galaxy, accounting for modeled losses due to stellar death and winds. This difference is important as mass losses can account for $>30\%$ of $M_{formed}$ (as measured from the CANDELS galaxies) and is heavily dependent on the age of a galaxy's stars. Therefore, using $M_{\rm formed}$ is crucial for measuring and comparing the stochasticity of star formation in galaxies with different stellar mass assembly scenarios and redshifts. Additionally, $t_{gal}$ and $t_{\rm univ}$ can be quite different, especially for a sample that includes many young galaxies (as we discussed in Section \ref{subsec:SFH_results}). Since the primary goal of our method is to understand star formation stochasticity as a function of redshift, the difficulty in reliably measuring $t_{gal}$ demands that $t_{\rm univ}$ prevail\edits{s} as the more direct quantity. 




\subsection{Star Formation Stochasticity Results}

\begin{figure*}
\begin{center}
\includegraphics[width=1\textwidth]{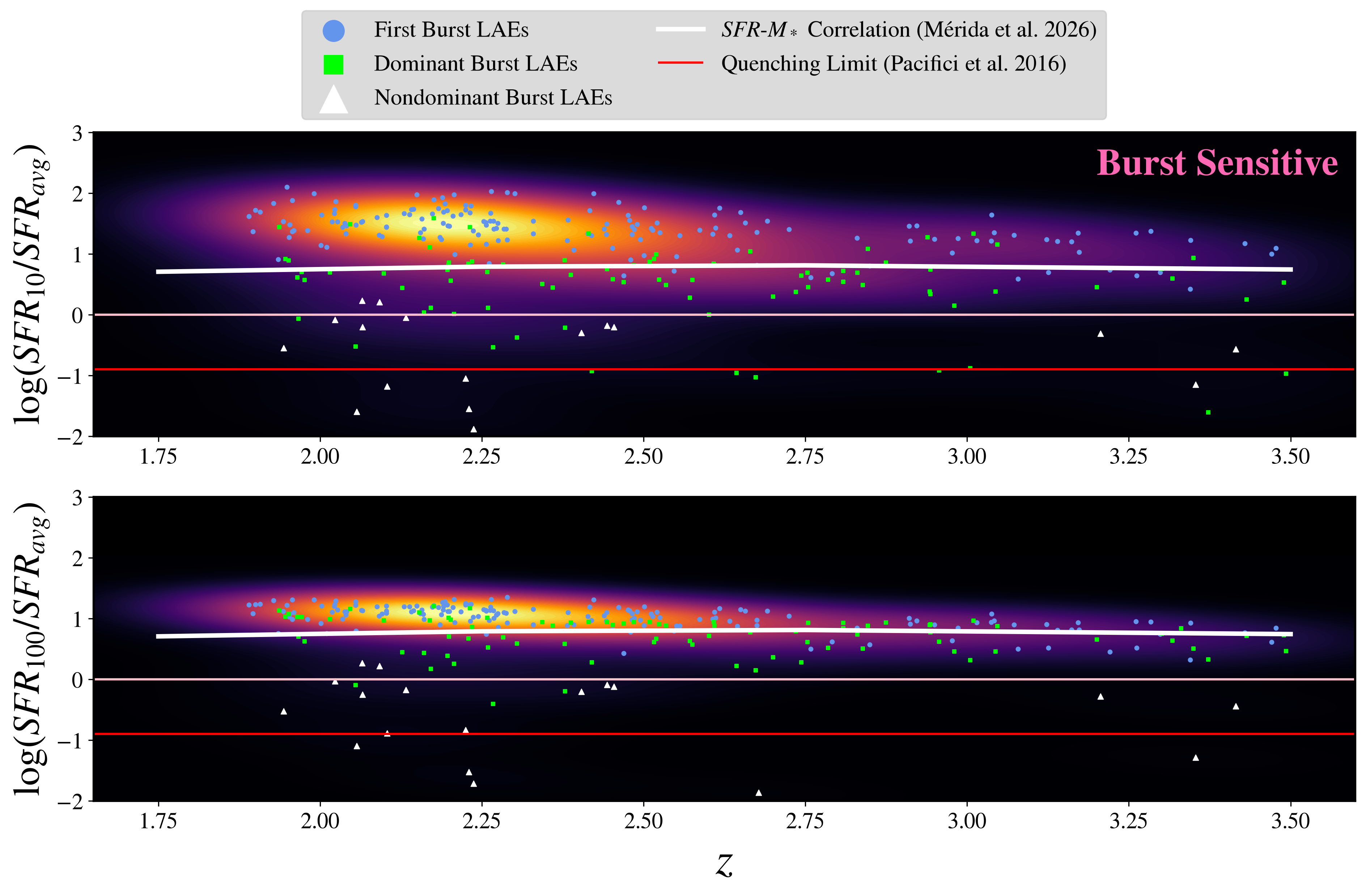}
\caption{Star Formation Stochasticity Diagrams. The $x$-axis represents the redshift ($z$) while the $y$-axis is the logarithmic ratio between the short-term and long-term average star formation rate. The top panel represents a burst-sensitive version of the diagram where the short term star formation rate is averaged over the last 10\,Myr; the bottom panel displays a generic version of the diagram where the short term star formation is averaged over the last 100\,Myr. The overall trend for the whole LAE sample is depicted with a heat map, while the individual galaxies are represented by points. These galaxies are identified by their LAE star formation archetypes, with First Burst LAEs shown as blue circles, Dominant Burst LAEs as green squares, and Nondominant Burst LAEs as white triangles. The SFR-M$_*$ correlation from \citet{merida2026probing} is presented as a white horizontal line, SFR = SFR$_{\rm avg}$ is a pink horizontal line, and the quenching limit from \citet{pacifici2016timing} is shown as a red horizontal line. 
} 
\label{fig:stochastiocity_diagram}
\end{center}
\end{figure*}

Figure \ref{fig:stochastiocity_diagram} presents two versions of the new Star Formation Stochasticity Diagram: a ``burst sensitive'' version employing $\log({\rm SFR}_{10}/{\rm SFR}_{\rm avg})$ and a generic version employing $\log({\rm SFR}_{100}/{\rm SFR}_{\rm avg})$. \edits{In order to compare with} a general population of star-forming galaxies on the Star Formation Stochasticity diagram, we convert the SFR-M$_*$ correlations from \citet{merida2026probing} using Equation \ref{eq:switch}. \edits{For the general population, we assume a mass retention fraction of $67$\%. This generalized retention fraction is derived from a comparable generic population of galaxies in the CANDELS dataset.} This conversion is performed for each redshift bin, taking $t_{\rm univ}$ from the bin's central redshift. We apply this procedure for both the generic and classic versions of the diagram. On average, SFR$_{10}$ and SFR$_{100}$ should be equal for non-bursting star-forming galaxies. For this reason, we use this correlation in both the generic and burst sensitive Star Formation Stochasticity Diagrams. 

First Burst LAEs lie above the median SFR-M$_*$ correlation, indicating star-bursting behavior. Dominant Burst LAEs lie near the median SFR-M$_*$ correlation, indicating generic star-forming behavior. Lastly, Nondominant Burst LAEs below the median correlation, indicating a more quiescent behavior. \edits{The Star Formation Stochasticity measurements for each archetype are presented in Table \ref{tab:stochasticity}.}
\begin{table}[h!]
\centering
\caption{Star Formation Stochasticity of LAEs by archetype.}\label{tab:stochasticity}
\textbf{Burst Sensitive: $\log({\rm SFR}_{10}/{\rm SFR}_{\rm avg})$}\\
\vspace{-0.2cm}
\begin{tabular}{lcc}
\hline\hline
Archetype & Median & Standard Deviation\\ 
\hline
First Burst & 1.42 & 0.33\\
Dominant Burst & 0.61 & 0.62\\
Nondominant Burst & -0.56 & 0.93\\
\hline
\vspace{0.2cm}

\end{tabular}
\textbf{Generic: $\log({\rm SFR}_{100}/{\rm SFR}_{\rm avg})$}\\
\vspace{-0.2cm}
\begin{tabular}{lcc}
\hline\hline
Archetype & Median & Standard Deviation\\ 
\hline
First Burst & 1.07 & 0.20\\
Dominant Burst & 0.77 & 0.30\\
Nondominant Burst & -0.48 & 0.86\\
\hline
\end{tabular}
\end{table}

To search for the presence of subpopulations within the LAE sample, we employ a Gaussian Mixture Model (GMM\null). A GMM is a machine learning algorithm used to decompose a dataset into a ``mixture'' of Gaussian components. Functionally, a GMM is a probabilistic model, i.e., it relies on sampling random combinations of possible Gaussian components, each with an assigned empirical probability. To select the optimal components from the sample of possible outcomes, we use the Bayesian Information Criterion (BIC), which maximizes likelihood while minimizing model complexity. 

\begin{figure*}
\begin{center}
\includegraphics[width=0.95\textwidth]{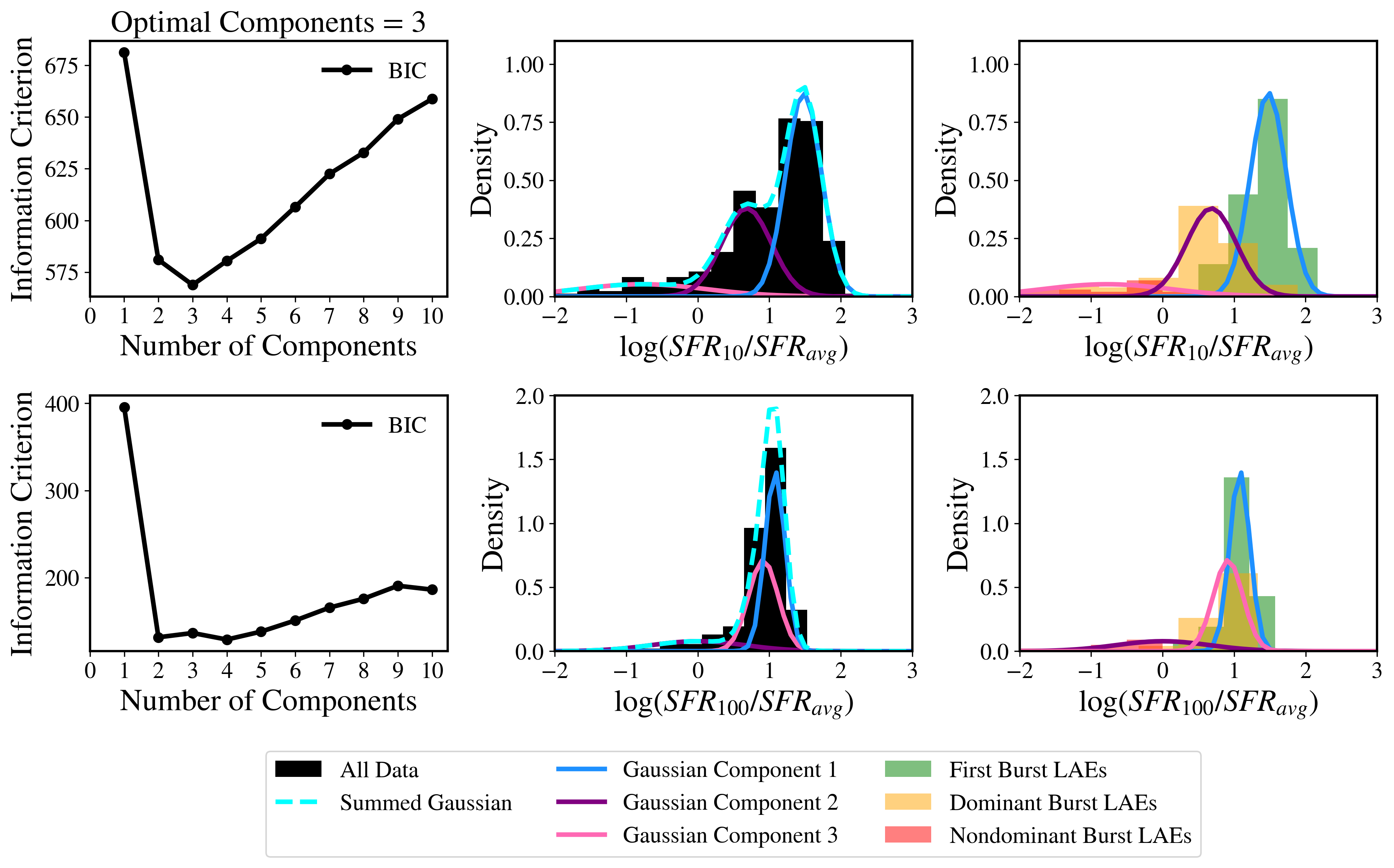}
\caption{Gaussian Mixture Model results for $\log({\rm SFR}_{10}/{\rm SFR}_{\rm avg})$ (top row) and $\log({\rm SFR}_{100}/{\rm SFR}_{\rm avg})$ (bottom row). The top left panel depicts the Bayesian Information Criterion for different numbers of Gaussian components, revealing that three components are preferred. The middle panels depict a histogram of all data in $\log({\rm SFR/SFR}_{\rm avg})$ (black) overlaid with the three Gaussian components (blue, pink, and purple, respectively) and the overall summed Gaussian (cyan). The right panels depict the same three Gaussian components overlaid with histograms for the First Burst (green), Dominant Burst (yellow), and Nondominant Burst (red) archetypes. This figure reveals that the three statistically motivated Gaussian components in $\log({\rm SFR}_{10}/{\rm SFR}_{\rm avg})$ space  recover the three LAE archetypes defined by their full star formation histories, with the separation clearer in the burst-sensitive analysis of the top row.} 
\label{fig:gaussian_mix}
\end{center}
\end{figure*}

For this analysis, we apply a GMM to our stochasticity metrics for the full sample across all redshifts. The top panel of Figure \ref{fig:gaussian_mix} represents the solution for the burst sensitive stochasticity metric $\log({\rm SFR}_{10}/{\rm SFR}_{\rm avg})$; the bottom panel represents the solution for the generic metric $\log({\rm SFR}_{100}/{\rm SFR}_{\rm avg})$. The leftmost panels of Figure \ref{fig:gaussian_mix} demonstrate the BIC for different numbers of Gaussian components. The BIC reveals that the burst sensitive metric has a significant preference for 3 components, while the generic metric is less well constrained. Since the generic metric does not show a strong preference between 2, 3, and 4 Gaussian components, \edits{we choose to use 3 components for direct comparison with the burst sensitive metric}. The middle panels of Figure \ref{fig:gaussian_mix}, presents histograms of $\log({\rm SFR/SFR}_{\rm avg})$ overlaid with the three Gaussian components and the overall summed Gaussian. The right panels of Figure \ref{fig:gaussian_mix} again presents the three Gaussian components, this time over the $\log({\rm SFR/SFR}_{\rm avg})$ histograms for the First Burst, Dominant Burst, and Nondominant burst archetypes. The Gaussian components clearly recover the empirical archetypes of \citet{firestone25}. This recovery is clearest in the burst sensitive version, which is best suited for understanding star-bursting galaxies such as LAEs. \edits{For the burst sensitive metric, the GMM component means are 1.48, 0.69, and -0.77; the covariances are 0.06, 0.12, and 0.80; and the weights are 0.55, 0.33, and 0.12.} The three LAE SFH archetypes are motivated and well-described by the $\log({\rm SFR}_{10}/{\rm SFR}_{\rm avg})$ metric. 



\section{Conclusions}\label{sec:conc}



In this study, we pair data from the HETDEX spectroscopic LAE survey with rest-UV-through-NIR photometry from CANDELS to produce a sample of 270 LAEs at $1.9 < z < 3.5$ in the COSMOS Field. Using the Dense Basis non-parametric Gaussian process-based SFH reconstruction code, we produced star formation histories for 270 LAEs and studied their star formation phase compared to generic star forming galaxies. 
The main conclusions of this work are summarized below. 

\begin{enumerate}[leftmargin=1.3\parindent]

    \item The spectroscopic HETDEX LAE sample yields star formation histories that agree with the three archetypes defined in \citet{firestone25}: \textit{First Burst} at the time of observation ($61\%$), \textit{Dominant Burst} showing the highest SFR at the time of observation but with at least one significant star formation episode in the past ($31\%$), and \textit{Nondominant Burst} describing the case where the highest SFR occurred in the past ($7\%$). This analysis results in a Formative Burst characterization for $\sim$92\% of the HETDEX LAE sample. 

    \item The traditional SFR-M$_*$ diagram utilizing SFR$_{100}$ does not adequately characterize the star-bursting behavior of First Burst LAEs. We suggest using a Burst Sensitive approach by comparing SFR$_{10}$ to stellar mass. 

    \item We introduce the Star Formation Stochasticity Diagram, a diagnostic tool that probes variations in galaxies' short-term versus long-term SFRs across cosmic time. 

    \item Using a burst sensitive Star Formation Stochasticity Diagram ($\log({\rm SFR}_{10}/{\rm SFR}_{\rm avg})$ vs.\ $z$), we compare the properties of LAEs to those of generic star forming galaxies as defined by the median correlation from \citet{merida2026probing}. First Burst LAEs lie above the median SFR-M$_*$ correlation, indicating star-bursting behavior, Dominant Burst LAEs lie near the median SFR-M$_*$, as expected for generic star-forming galaxies, and Nondominant Burst LAEs fall below the median correlation, indicating a more quiescent behavior. The archetypes are less well-separated  in a generic Star Formation Stochasticity Diagram ($\log({\rm SFR}_{100}/{\rm SFR}_{\rm avg})$ vs.\ $z$). 

    \item A component decomposition of the distribution of $\log({\rm SFR}_{10}/{\rm SFR}_{\rm avg})$ using a Gaussian Mixture Model reveals three subpopulations of LAEs. These \edits{independently defined} subpopulations are remarkably consistent with the archetypes defined using their full star formation histories, implying that they are robust and statistically significant. We can, therefore, summarize the overall stellar mass assembly of LAEs with this metric, even without viewing their full star formation histories.
    
\end{enumerate}

We have demonstrated that it is possible to characterize the star formation stochasticity of LAEs spanning a broad redshift range ($\Delta z \sim 1.6$). With the advantage of smooth non-parametric star formation histories, we are able to accurately assess the behavior of LAEs with an acknowledgment of their diverse formation histories. With this work, we are one step closer to characterizing the complexity within these galaxies that probe galaxy formation, galaxy evolution, cosmology, and reionization in the high-$z$ \edits{U}niverse. 

\section{Acknowledgments}



HETDEX is led by the University of Texas at Austin McDonald Observatory and Department of Astronomy with participation from the Ludwig-Maximilians-Universit\"at M\"unchen, Max-Planck-Institut f\"ur Extraterrestrische Physik (MPE), Leibniz-Institut f\"ur Astrophysik Potsdam (AIP), Texas A\&M University, Pennsylvania State University, Institut f\"ur Astrophysik G\"ottingen, The University of Oxford, Max-Planck-Institut f\"ur Astrophysik (MPA), The University of Tokyo and Missouri University of Science and Technology.

Observations for HETDEX were obtained with the Hobby-Eberly Telescope (HET), which is a joint project of the University of Texas at Austin, the Pennsylvania State University, Ludwig-Maximilians-Universit\"at M\"unchen, and Georg-August-Universit\"at G\"ottingen. The HET is named in honor of its principal benefactors, William P. Hobby and Robert E. Eberly. The Visible Integral-field Replicable Unit Spectrograph (VIRUS) was used for HETDEX observations. VIRUS is a joint project of the University of Texas at Austin, Leibniz-Institut f\"ur Astrophysik Potsdam (AIP), Texas A\&M University, Max-Planck-Institut f\"urExtraterrestrische Physik (MPE), Ludwig-Maximilians-Universit\"at M\"unchen, Pennsylvania State University, Institut f\"ur Astrophysik Göttingen, University of Oxford, and the Max-Planck-Institut fur Astrophysik (MPA).

The authors acknowledge the Texas Advanced Computing Center (TACC)\footnote{See http://www.tacc.utexas.edu for information about the Texas Advanced Computing Center.} at The University of Texas at Austin for providing high performance computing, visualization, and storage resources that have contributed to the research results reported within this paper. 

Funding for HETDEX has been provided by the partner institutions, the National Science Foundation, the State of Texas, the US Air Force, and by generous support from private individuals and foundations.

This material is based upon work supported by the NSF Graduate Research Fellowship Program under Grant No. DGE-2233066 to NF. NF and EG acknowledge support from NSF grant AST-2206222 and DOE grant DE-SC0010008. NF, EG, and CO acknowledge support from NASA Astrophysics Data Analysis Program grant 80NSSC22K0487. NF would like to thank the LSST-DA Data Science Fellowship Program, which is funded by LSST Discovery Alliance, NSF Cybertraining Grant 1829740, the Brinson Foundation, and the Moore Foundation; her participation in the program has benefited this work greatly. CO is supported by an LSST-DA Catalyst Fellowship funded through Grant 62192 from the John Templeton Foundation to LSST Discovery Alliance. RC and CG acknowledge support from the National Science Foundation under grant AST-2408358. SS acknowledges support from the National Science Foundation under grants NSF-2219212 and NSF-2511145. 
\edits{We thank the anonymous referee for sharing their expertise and thoughtful suggestions, which have improved this work.}

\bibliography{sample631}{}
\bibliographystyle{aasjournal}

\end{document}